\documentstyle[prl,aps,twocolumn,amsmath,graphicx]{revtex} 

\begin{document}

\title{Phase diagram and magnons in quasi-one-dimensional dipolar
anti\-ferro\-mag\-nets}

\author{M. Hummel and F. Schwabl}

\address{Physik-Department, Technische Universit\"at M\"unchen, D-85747
  Garching, Germany}

\author{C. Pich}

\address{Physics Department, University of California, Santa Cruz, CA 95064}

\date{\today}

\maketitle

\begin{abstract}
  We investigate antiferromagnetic spin chains, which are coupled by a weak
  antiferromagnetic exchange interaction. The spins are located on a hexagonal
  lattice, i.e. frustration is present when three-dimensional order sets in.
  Typical realizations of such systems are the halides ABX$_3$. In this work we
  particularly study the role of the long-range dipolar interaction within the
  framework of a Heisenberg model with nearest-neighbor exchange and additional
  dipolar interaction.  We perform a classical ground-state analysis and show
  that the spin configuration is sensitively dependent on $\kappa'$, the ratio
  of the dipolar interaction to the interchain inter\-action, as a consequence
  of their competing character. Several commensurate and incommensurate phases
  arise in the different regions of the parameter space.  The ground-state
  investigations are supplemented by a stability analysis by means of a linear
  spin-wave calculation. From the magnon spectra we can show that all
  commensurate phases are stable against fluctuations. In comparison with
  experiments (CsMnBr$_3$, RbMnBr$_3$) we obtain good agreement for the energy
  gaps. From this we conclude that the dipolar interaction is the most
  important source of anisotropy in these Mn-compounds.
\end{abstract}


\section{Introduction}
\label{sec:intro}
  \noindent
  Unconventional magnetic systems have attracted the interest of experimental
  and theoretical physicists in the last few years\cite{Sherrington}. In these
  systems competing interactions and/or geometric frustration due to the
  underlying lattice can lead to unconventional ground states, magnon spectra
  and magnetic phase diagrams\cite{diep:94}. Furthermore, fluctuations are
  enhanced in systems with frustrated ground states as well as in low
  dimensions.
  
  An interaction, which is often competing with respect to the exchange
  interaction is the dipole-dipole interaction\cite{pich97}.  In real systems
  the dipole-dipole interaction (DDI) is always present in addition to the
  short-range exchange interaction. Although its energy is lower than the
  exchange energy it plays an important role in low-dimensional systems due to
  its anisotropic and long-range character.
  
  The most famous quasi-one-dimensional systems are the ternary compounds
  ABX$_3$ (A alkaline, B transition metal and X halogen), which have been
  studied intensively theoretically and experimentally in the context of
  Haldane's phase \cite{Haldane} and solitonic excitations \cite{Steiner91}. In
  these systems the carrier of the magnetic moment, the B-ions, are located on
  a hexagonal lattice\cite{Collins:98}. In this work, the Mn compounds are of
  particular interest.  Because the angular momentum $L$ is zero, no
  crystal-field splitting occurs in these systems and the dipole-dipole
  interaction should be the most important anisotropy.
  
  The influence of the DDI in quasi-one-dimensional, antiferromagnetic
  spin chain systems has not yet been studied very thoroughly. Instead, the DDI
  is often replaced by a single-ion anisotropy\cite{kadowaki} or the coupling
  between spin chains is neglected\cite{dietz}.

\section{Model} 
\label{sec:model}
  \noindent
  The Hamiltonian of the dipolar antiferromagnet reads
  \begin{equation}
    H=-\sum_{l \neq l'} \sum_{\alpha \beta} \bigl( J_{ll'} \delta_{\alpha 
    \beta} + A_{ll'}^{\alpha\beta} \bigr) S_l^{\alpha} S_{l'}^{\beta} \; ,
    \label{Hamiltonian}
  \end{equation}
  with spins ${\bf{S}}_l$ at hexagonal lattice sites ${\bf{x}}_l$.  The first
  term describes the exchange interaction $J_{ll'}$ which includes the
  intrachain as well as the interchain interaction.  In the following we
  consider only nearest-neighbor exchange, i.e.
  \begin{displaymath}
    J_{l l'} = \left\{ 
      \begin{array}{lll}
        -J & l,l' & \mbox{along the chains}\\
        -J' &  l,l' & \mbox{within the basal plane} 
      \end{array}
    \right. \quad .
  \end{displaymath}  
  For a hexagonal lattice the Fourier trans\-form of the exchange energy is
  given by
  \begin{equation}
    J_{\bf{q}}=-2J \cos q_z - 2J' \Bigl( 
    \cos q_x + 2 \cos \bigl( \frac{q_x}{2} \bigr) \cos \bigl( 
    \frac{\sqrt{3}}{2} q_y \bigr) \Bigr) \; .
  \end{equation}
  Here and in the following we measure wave vectors in chain direction in units
  of $1/c$, and wave vectors within the basal planes in units of $1/a$.  The
  second term in Eq.~(\ref{Hamiltonian}) is the classical
  dipole-dipole interaction
  \begin{eqnarray}
    A_{ll'}^{\alpha\beta}= -\frac{1}{2} {(g \mu_B)^2} && \Biggl(
    \frac{\delta_{\alpha\beta}}{{|{\bf{x}}_l-{\bf{x}}_{l'}|}^3}\nonumber\\
    &&- \frac{3 ({{\bf{x}}_l-{\bf{x}}_{l'})}_{\alpha} 
      {({\bf{x}}_l-{\bf{x}}_{l'})}_{\beta}}{{|{\bf{x}}_l-{\bf{x}}_{l'}|}^5}
    \Biggr) \, .
    \label{ddw}
  \end{eqnarray}  
  This term is evaluated by means of the Ewald summation
  technique\cite{Ewald,Maradudin}, which allows the consideration of the
  long-range nature of the three-dimensional DDI in terms of fast convergent
  sums.

\section{Ground states} 
\label{sec:gs} 
\begin{figure}
  \narrowtext
    \begin{center}
      \includegraphics[width=0.3\textwidth]{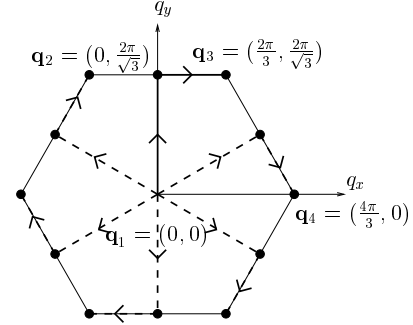}
        \caption{\label{figbz}Paths of the wave vector that mini\-mize the 
        ground-state energy in the Brillouin zone of the hexagonal lattice
        ($q_z=\pi$).}
    \end{center} 
  \end{figure}

\begin{figure}
  \narrowtext
    \begin{center}
      \includegraphics[width=0.47\textwidth]{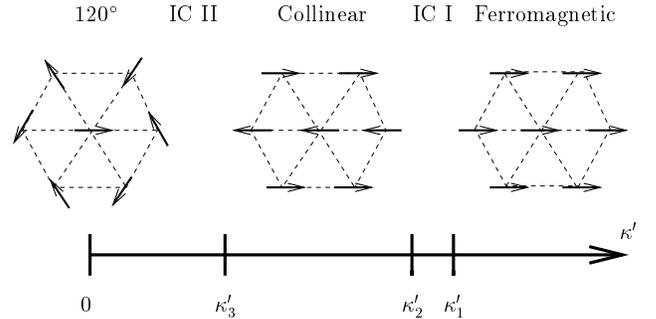}
      \caption{\label{fig2}Spin configurations within the basal plane for 
        dif\-fer\-ent ratios of dipolar to interchain interaction $\kappa'$.
        Only for $\kappa'=0$ the 120$^\circ$-structure is established. For
        infinitesimal $\kappa'$ the phase IC II is favored. The spin
        configurations of incommensurate phases are not sketched.}
    \end{center} 
  \end{figure}

  \noindent
  In this section we calculate the classical ground states of the system as
  a function of the ratio of the dipolar energy to the interchain exchange
  interaction
  \begin{equation}
    \kappa' = \frac{(g\mu_B)^2}{V_z J'}  \; ,
  \end{equation}
  where $V_z=\frac{\sqrt{3}}{2}a^2 c$ is the volume of the primitive cell. Due
  to the quasi-one-dimensionality of the systems (the ratio $J'/J$ is $10^{-2}
  \ldots 10^{-3}$) we consider only antiferromagnetic spin configurations along
  the chain axis, i.e., we restrict $q_z$ to $\pi$. The ground-state energy
  reads
  \begin{equation}
    E_g = - \sum_{l \neq l'} S_l M_{l l'} S_{l'} 
    = - \sum_{\bf{q}} S_{\bf{q}} M_{\bf{q}} S_{-\bf{q}}
  \end{equation}
  with
  \begin{equation}
    M_{\bf{q}}=\begin{pmatrix}
      J_{\bf q} + A_{\bf q}^{11} & A_{\bf q}^{12} & A_{\bf q}^{13}\\
      A_{\bf q}^{12} & J_{\bf q} + A_{\bf q}^{22} & A_{\bf q}^{23}\\
      A_{\bf q}^{13} & A_{\bf q}^{23} & J_{\bf q} + A_{\bf q}^{33} \\
    \end{pmatrix}
    \; ,
    \label{matrix}
  \end{equation}  
  where $A_{\bf q}^{13} = A_{\bf q}^{23} = 0$ for $q_z=\pi$. The ground states
  are specified by those wave vectors which belong to the largest eigenvalues
  of the matrix $M_{\bf{q}}$ in Eq.~(\ref{matrix}).  We obtain the following
  phases for decreasing $\kappa'$ (for the lattice constants we took the values
  of CsMnBr$_3$ \cite{goodyear}):

  \begin{enumerate}
    \item Ferromagnetic phase: $\kappa' > \kappa_1' = 200.50$
      
      In the region where the dipolar energy is large compared to the
      interchain exchange, the minimum of the ground-state energy is reached at
      ${\bf q}_1 = (0,0,\pi)$. This means that spins within basal planes are
      ordered ferromagnetically, but still antiferromagnetically along the
      chains.
      
    \item Incommensurate phase I: $200.12 = \kappa_2' < \kappa' < \kappa_1'$

      The ground state is an incommensurate phase in this parameter region.
      The wave vector moves continuously from ${\bf q}_1$ to the wave vector 
      ${\bf q}_2 = (0,\frac{2\pi}{\sqrt{3}},\pi)$ (thick solid line in 
      Fig.~\ref{figbz}) or any other path rotated by 60$^\circ$ 
      (dashed lines in Fig.~\ref{figbz}).

    \item Collinear phase: $17.25 = \kappa_3' < \kappa' < \kappa_2'$
      
      Spins within basal planes are oriented ferromagnetically in chains, 
      that are aligned antiferromagnetically to one another (see
      Fig.~\ref{fig2}). Because of the six-fold symmetry, there are six
      such ground states resulting from rotation of the ferromagnetic chains.
      Note that the continuous degeneracy is lifted.

    \item Incommensurate phase II: $0 < \kappa' < \kappa_3'$
      
      In this incommensurate phase the wave vector mo\-ves from ${\bf q}_2$ to
      ${\bf q}_3 = (\frac{2 \pi}{3},\frac{2\pi}{\sqrt{3}},\pi)$. The
      incommensurability appears because of the finite slope of the dipolar
      tensor at ${\bf q}_3$ \cite{Shiba} and the parabolic behavior of the
      exchange energy.

     \item 120$^\circ$-structure: $\kappa' = 0$

      This ground state is characterized by a three-sublattice spin
      configuration in each basal plane.

  \end{enumerate}  

  \noindent
  The coupling of the spin space to the real space induced by the DDI forces
  the spins to align within the lattice basal planes for all four phases,
  in which the DDI is non-zero. Thus, the DDI leads to an in-plane anisotropy.

  \noindent
  The phase diagram for the whole parameter region of $\kappa'$ is shown in
  Fig. \ref{fig2}.
 
  \noindent
  In summary, we find three commensurate and two incommensurate phases for
  arbitrary value of $\kappa'$.

\section{Magnon spectra} 
\label{sec:ms} 
  \noindent
  The spin-wave calculation is of interest in its own right and also serves to
  scrutinize the stability of the phases found against fluctuations. To that
  end we write the Hamiltonian (\ref{Hamiltonian}) in terms of creation and
  annihilation operators employing the Holstein-Primakoff
  transformation~\cite{HP}. For low temperatures an expansion up to bilinear
  terms can be used, leading to linear spin-wave theory. The Hamiltonian
  then is diagonalized with a Bogoliubov transformation from which we obtain
  the spin-wave frequencies. This investigation is restricted to the
  commensurate phases, because the infinite primitive cell of the
  incommensurate phases resists such an analysis.
  
\noindent
{\bf Ferromagnetic phase}

  The magnon spectrum of the ferromagnetic phase is stable for the parameter
  region given in Sec.~\ref{sec:gs}. The rotational invariance of the spins 
  around the chain axis leads to a Goldstone mode in the spectrum\cite{hummel}.
\noindent
{\bf Collinear phase}

  The collinear phase of Sec.~\ref{sec:gs} is also stable against
  fluctuations. There is no Goldstone mode due to the discrete 
  degeneracy of the ground state. 

\noindent
{\bf 120\boldmath$^\circ$-structure}

  We argued in Sec.~\ref{sec:gs} that the 120$^\circ$-structure is unstable for
  infinitesimal dipolar energy due to a linear slope of the dipolar tensor at 
  the ordering wave vector. However, we performed a spin-wave calculation based
  on a commensurate 120$^\circ$-structure, where the spins are located within
  the basal planes of the lattice\cite{oyedele}. It turns out that the
  spin-wave spectrum is stable for weak dipolar energies, from which we 
  conclude that this commensurate ground state is a good approximation. 

  \begin{figure}
  \narrowtext
    \begin{center}
      \includegraphics[width=0.39\textwidth]{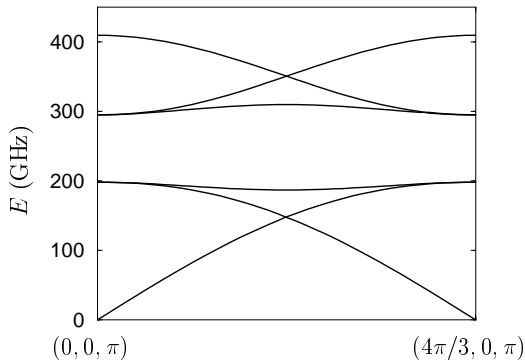}
      \caption{\label{figspectrum}Magnon spectrum of the 120$^\circ$-structure
      perpendicular to the chain direction.}
    \end{center} 
  \end{figure}

  \noindent
  The spin-wave frequencies resulting for the dipolar 120$^\circ$-structure for
  CsMnBr$_3$ are shown in Fig.~\ref{figspectrum}, where we used $J=215$ GHz and
  $J'=0.41$ GHz respectively\cite{falk}. This leads to $\kappa' = 0.774$, i.e.,
  this substance is in the IC II region of the phase diagram in
  Fig.~\ref{fig2}.  Including the DDI, from the three Goldstone modes only one
  survives reflecting the unchanged rotational symmetry around the chain axis.
  
  The spin-wave gaps at ${\bf q} = 0$ amount to $E_{0,1} = 198$~GHz, $E_{0,2} =
  295$ GHz and $E_{0,3} = 410$ GHz for CsMnBr$_3$, which compares favorably
  with the experimental values \cite{falk,gaulin}.  Note that this calculation
  has no free parameter to fit, since the dipolar energy is determined by the
  lattice constants. We also calculated the spin-wave gaps for RbMnBr$_3$;
  neglecting crystal distortions we also obtain good agreement with the
  experiment\cite{heller}.
  
  Thus, we do not need any single-ion anisotropy to explain these results.

  \section{Summary}  
  \noindent
  We found three commensurate and two incommensurate phases for different
  values of the ratio of dipolar to interchain interaction due to the competing
  character of those two interactions. We showed via linear spin-wave theory
  that all commensurate phases are stable against fluctuations and that the
  incommensurate phase IC II can be approximated by a 120$^\circ$-structure
  for weak dipolar energies.
  
  The spin-wave gaps of CsMnBr$_3$ and RbMnBr$_3$ are in good agreement with
  the experiment, which shows that the dipolar energy is the most important
  source of anisotropy in these Mn-compounds.

  This work has been supported by the BMBF under contract number 03-SC5-TUM 0 
  and the DFG under contract number PI 337/1-2.

\end{document}